\documentclass{aastex61}

\submitjournal{ApJ}

\shorttitle{Dust condensations in the Orion Bar}
\shortauthors{Qiu et al.}

\begin{document}

\title{Detection of dust condensations in the Orion Bar Photon-dominated Region}

\correspondingauthor{Keping Qiu}
\email{kpqiu@nju.edu.cn}

\author[0000-0002-5093-5088]{Keping Qiu}
\affil{School of Astronomy and Space Science, Nanjing University, 163 Xianlin Avenue, Nanjing 210023, China}
\affil{Key Laboratory of Modern Astronomy and Astrophysics (Nanjing University), Ministry of Education, Nanjing 210023, China}

\author{Zeqiang Xie}
\affil{School of Astronomy and Space Science, Nanjing University, 163 Xianlin Avenue, Nanjing 210023, China}
\affil{Key Laboratory of Modern Astronomy and Astrophysics (Nanjing University), Ministry of Education, Nanjing 210023, China}

\author{Qizhou Zhang}
\affiliation{Harvard-Smithsonian Center for Astrophysics, 60 Garden Street, Cambridge, MA 02138, U.S.A.}



\begin{abstract}

\quad We report Submillimeter Array dust continuum and molecular spectral line observations toward the Orion Bar photon-dominated region (PDR). The 1.2~mm continuum map reveals, for the first time, a total of 9 compact ($r<0.01$~pc) dust condensations located within a distance of $\sim$0.03~pc from the dissociation front of the PDR. Part of the dust condensations are seen in spectral line emissions of CS~(5--4) and H$_{2}$CS~($7_{1,7}$--$6_{1,6}$), though the CS map also reveals dense gas further away from the dissociation front. We detect compact emissions in H$_{2}$CS ($6_{0,6}$--$5_{0,5}$), ($6_{2,4}$--$5_{2,3}$) and C$^{34}$S, C$^{33}$S (4--3) toward bright dust condensations. The line ratio of H$_{2}$CS ($6_{0,6}$--$5_{0,5}$)/($6_{2,4}$--$5_{2,3}$) suggests a temperature of $73\pm58$~K. A non-thermal velocity dispersion of $\sim$0.25--0.50~km\,s$^{-1}$ is derived from the high spectral resolution C$^{34}$S data, and indicates a subsonic to transonic turbulence in the condensations. The masses of the condensations are estimated from the dust emission, and range from 0.03 to 0.3~$M_{\odot}$, all significantly lower than any critical mass that is required for self-gravity to play a crucial role. Thus the condensations are not gravitationally bound, and could not collapse to form stars. In cooperating with recent high resolution observations of the surface layers of the molecular cloud in the Bar, we speculate that the condensations are produced as a high-pressure wave induced by the expansion of the H{\scriptsize II} region compresses and enters the cloud. A velocity gradient along a direction perpendicular to the major axis of the Bar is seen in H$_{2}$CS~($7_{1,7}$--$6_{1,6}$), and is consistent with the scenario that the molecular gas behind the dissociation front is being compressed.

\end{abstract}




\section{Introduction} \label{sec:intro}
Young massive OB stars produce strong ultraviolet radiation that significantly influences the structure, chemistry, thermal balance, and dynamical evolution of the nearby interstellar medium \citep{Hollenbach97}. Their extreme-ultraviolet (EUV) photons ionize the surrounding gas and form H{\scriptsize II} regions. Photon-dominated or photo-dissociated regions (PDRs) start at the H{\scriptsize II} region/neutral cloud boundary where the EUV radiation vanishes and the far-ultraviolet (FUV) radiation becomes dominant, dissociating H$_2$ and ionizing heavier elements. The gas inside the PDR transits from atomic to molecular as the FUV flux decreases due to dust extinction and H$_2$ absorption \citep[e.g.,][]{Goicoechea16}. PDRs exist in many astrophysical environments and span a board range of spatial scales, from the nuclei of starburst galaxies \citep[e.g.,][]{Fuente08} to the illuminated surfaces of protoplanetary disks \citep[e.g.,][]{Agundez08}. The study of PDRs could also help to understand the impact from massive young stars on subsequent star formation in nearby molecular clouds.

The Orion Bar is probably the best studied PDR in our Galaxy. It is located between the Orion Molecular Cloud 1 and the H{\scriptsize II} region excited by the Trapezium cluster, and is exposed to a FUV field a few $10^4$ times the mean interstellar radiation field. Owing to its proximity \citep[417 pc,][]{Menten07} and nearly edge-on orientation, the Bar provides an ideal laboratory for testing PDR models \citep[e.g.,][]{Jansen95,Gorti02,Andree17} and a primary target for observational studies of physical and chemical structures of PDRs \citep[e.g.,][]{Tielens93,Walmsley00,van09,Arab12,Peng12,Goicoechea16,Nagy17}. Observations of various molecular spectral lines have shown that the emissions could be better interpreted with an inhomogeneous density structure containing an extended and relatively low density ($n_{\rm H}\sim10^4$ -- $10^5$ cm$^{-3}$) medium and a compact and high density ($n_{\rm H}\sim10^5$ -- $10^6$ cm$^{-3}$) component \citep[e.g.,][]{Hogerheijde95,Young00,Leurini06,Leurini10,Goicoechea16}. However, due to the scarce of high resolution observations capable of spatially resolving the density structure, the nature of the high density clumps or condensations is still not well understood. \citet{Lis03} mapped the Bar in H$^{13}$CN (1--0) with the Plateau de Bure Interferometer (PdBI) at an angular resolution of about $5''$, and detected 10 dense clumps. They proposed that the H$^{13}$CN clumps are in virial equilibrium and may be collapsing to form stars. \citet{Goicoechea16} performed Atacama Large Millimeter/submillimeter Array (ALMA) HCO$^+$ (4--3) observations of the Bar and detected over-dense substructures close to the cloud edge, and found that the substructures have masses much lower than the mass needed to make them gravitationally unstable. These two interferometric observations both target molecular spectral lines. A high resolution map of the dust continuum emission of the Bar, which is highly desirable in constraining the mass and density of the dense condensations, is still lacking. Here we report our Submillimeter Array (SMA) observations of the dust continuum and molecular spectral line observations of the Bar. 

\section{Observations} \label{sec:obs}
The SMA\footnote{The SMA is a joint project between the Smithsonian Astrophysical Observatory and the Academia Sinica Institute of Astronomy and Astrophysics, and is funded by the Smithsonian Institution and the Academia
Sinica.} observations were carried out in 2009 and 2012. The observations in 2009 were taken on January 14th, January 30th, and February 3rd, with 5, 7, and 7 antennas, respectively, in the Sub-compact configuration. The weather conditions were good, with the zenith atmospheric opacity at 225 GHz, $\tau_{\rm 225GHz}$, in the range of 0.05 to 0.15. We observed two fields, one in the northeast (NE) centered at (R.A., decl.)$_{\rm J2000}$ = ($5^{\rm h}35^{\rm m}25.\!^{\rm s}2$, $-5{\arcdeg}24{\arcmin}34.\!{\arcsec}6$) and the other in the southwest (SW) centered at (R.A., decl.)$_{\rm J2000}$ = ($5^{\rm h}35^{\rm m}22.\!^{\rm s}1$, $-5{\arcdeg}25{\arcmin}13.\!{\arcsec}4$), and used Titan for flux calibration, Uranus and 3C454.3 for bandpass calibration, and J0423-013, J0607-085 for time dependent gain calibration. The 230~GHz receivers were tuned to cover rest frequencies of 234.76 -- 236.76 GHz in the lower sideband (LSB) and 244.76 -- 246.76 GHz in the upper sideband (USB). Signals from each sideband were processed by correlators consisting of 24 chunks with each chunk having a bandwidth of 104~MHz divided into 256~channels, resulting in a uniform spectral resolution of 406.25 kHz ($\sim$0.5~km\,s$^{-1}$). With this setup we could simultaneously observe the 1.2~mm continuum and molecular spectral lines of CS~(5--4) and H$_2$CS (7$_{1,6}$--6$_{1,6}$). Motived by the detection of dust continuum and spectral line emissions with the 2009 observations, we performed additional observations in 2012 to further constrain the physical conditions of the dense gas in the Bar. The observations were performed on 2012 January 1st with 7 antennas in the Compact configuration. The weather was good, with $\tau_{\rm 225GHz}$ varying from 0.07 to 0.18. We observed Callisto for flux calibration, 3C279 for bandpass calibration, and J0423-013, J0607-085 for time dependent gain calibration. By the time of our 2012 observations, the SMA correlators had been upgraded to be able to process signals across a bandwidth of 4~GHz divided into 48 chunks in each sideband. The 230~GHz receivers were then tuned to cover 192.26 -- 196.26 GHz in the LSB and 204.26 -- 208.26 GHz in the USB, and the correlators were configured to provide a spectral resolution of 812.5 kHz ($\sim$1.2~km\,s$^{-1}$) across all the chunks expect chunk 42, which covered C$^{34}$S (4--3) and was set to provide a high spectral resolution of 203.125 kHz ($\sim$0.3 km\,s$^{-1}$). This setup also covered spectral lines of CS~(4--3), C$^{33}$S (4--3), and H$_2$CS (6$_{0,6}$--5$_{0,5}$), (6$_{2,4}$--5$_{2,3}$). 

The raw data were calibrated using the IDL MIR package\footnote{\url{https://www.cfa.harvard.edu/~cqi/mircook.html}} and the calibrated visibilities were exported to MIRIAD for further processing. The visibilities were separated into continuum and spectral line data before imaging. Given that the frequency setups between the 2009 and 2012 observations differed by about 40.5~GHz and that the 2009 observations had a better $(u,v)$ coverage, we made the 1.2 mm continuum map with the 2009 data and obtained a synthesized beam with a full-width-half-maximum (FWHM) size of $3.\!''6{\times}3.\!''0$ and a position angle (PA) of $9^{\circ}$. The root mean square (RMS) noise level of the continuum map is 3.0~mJy\,beam$^{-1}$. Maps of the CS~(5--4) and H$_2$CS (7$_{1,6}$--6$_{1,6}$) lines, covered by the 2009 observations, have synthesized beams approximately the same as that of the 1.2~mm continuum map. The RMS noise level of the CS~(5--4) map is about 40~mJy\,beam$^{-1}$ per 0.5~km\,s$^{-1}$, while the H$_2$SC line is close to an atmospherical absorption feature and thus has a higher RMS noise level of 60~mJy\,beam$^{-1}$ per 0.5~km\,s$^{-1}$. Maps of the CS, C$^{33}$S, and C$^{34}$S (4--3), SO~(5$_4$--4$_3$), and H$_2$CS (6$_{0,6}$--5$_{0,5}$), (6$_{2,4}$--5$_{2,3}$) were made from the 2012 data. The CS, C$^{33}$S, C$^{34}$S, and SO maps have synthesized beams with a FWHM size about $4.\!''0{\times}2.\!''7$ and a PA of $-50^{\circ}$. The RMS noise levels are about 60~mJy\,beam$^{-1}$ per 1.2~km\,s$^{-1}$ for the CS, C$^{33}$S, and SO maps, and 120~mJy\,beam$^{-1}$ per 0.3~km\,s$^{-1}$ for the C$^{34}$S map. To optimize the signal-to-noise (S/N) ratios, a Gaussian taper of $4''\times4''$ were applied to the H$_2$CS lines during the imaging process, resulting in a synthesized beam with a FWHM size of $5.\!''0{\times}4.\!''5$ and a PA of $-71^{\circ}$, and an RMS noise level of about 50~mJy\,beam$^{-1}$ per 1.2~km\,s$^{-1}$. 

The \emph{Spitzer} IRAC data were obtained from the Spitzer archive (PID: 8334669). We adopted Post Basic Calibrated Data provided by the Spitzer Science Center.

\section{Results} \label{sec:results}

\subsection{Dust Continuum Emission} \label{subsec:cont}
\begin{figure}
\epsscale{1}
\plottwo{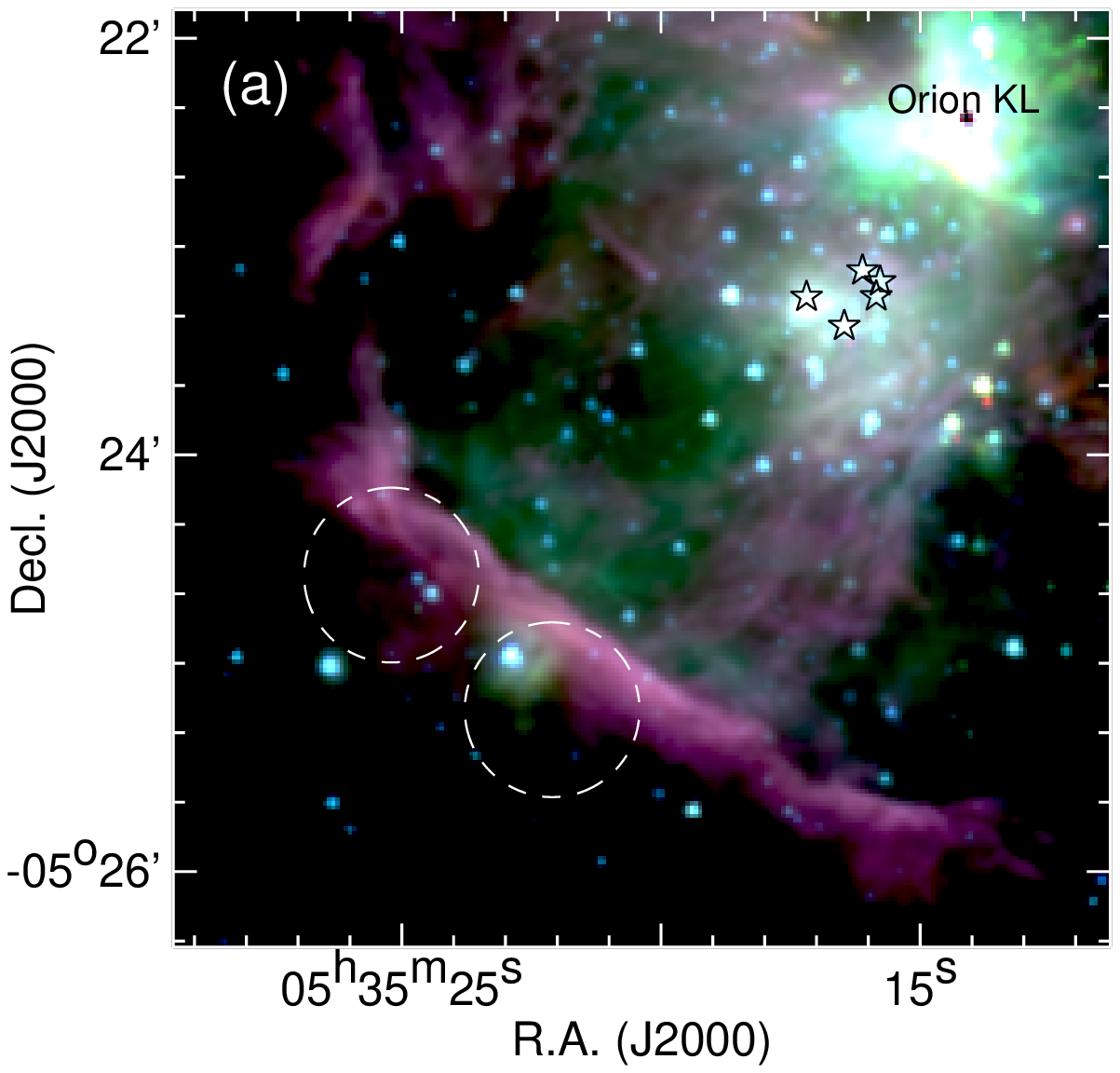}{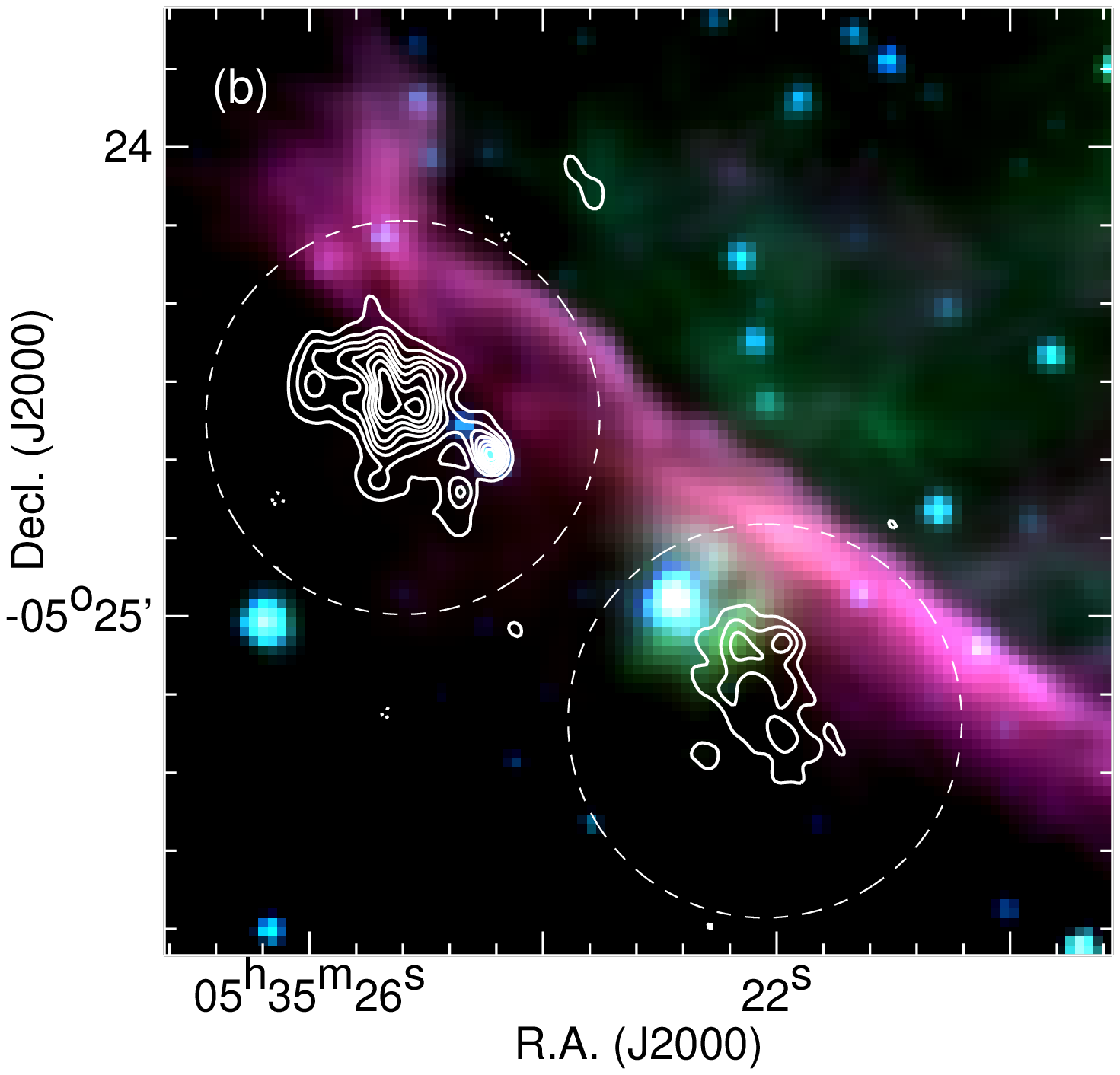}
\caption{(a) \emph{Spitzer} IRAC color-composite image with the 3.6, 4.5, and 5.8~$\mu$m emission coded in blue, green, and red, respectively. The bright Orion KL nebula in the northwest corner of the image has been marked. Five star symbols denote stars $\theta^1$ Orionis A--E in the Trapezium cluster. The Orion Bar is prominent in the 5.8~$\mu$m emission. Two white circles represent the fields of view of the SMA observations. (b) Close-up of the IRAC image focusing on the Orion Bar, overlaid with the contour map of the 1.2~mm continuum emission observed with the SMA. Contour levels start at and continue in steps of $\pm3\sigma$, where $\sigma=3$~mJy\,beam$^{-1}$ is the RMS noise level of the emission. For contour maps hereafter, positive emission is shown in solid (or filled) contours, and negative emission is shown in dotted contours with absolute contour levels the same as those for positive emission.
\label{fig:spitzer}}
\end{figure}

Figure~\ref{fig:spitzer}(a) shows a \emph{Spitzer} IRAC color-composite image of a $\sim4'\times4'$ region covering the Kleinmann-Low (KL) nebula, Trapezium cluster, and Orion Bar from northwest to southeast. The Bar stands out by virtue of its bright 5.8~$\mu$m emission, which is most likely FUV excited polycyclic aromatic hydrocarbon (PAH) emission and delineates an atomic layer \citep{Goicoechea16}. In Figure ~\ref{fig:spitzer}(b), the SMA 1.2~mm continuum observations reveal dust emission immediately behind the PAH ridge of the Bar, providing unambiguous evidence for the existence of dense and clumpy molecular gas shielded from FUV emission. 

\begin{figure}
\plotone{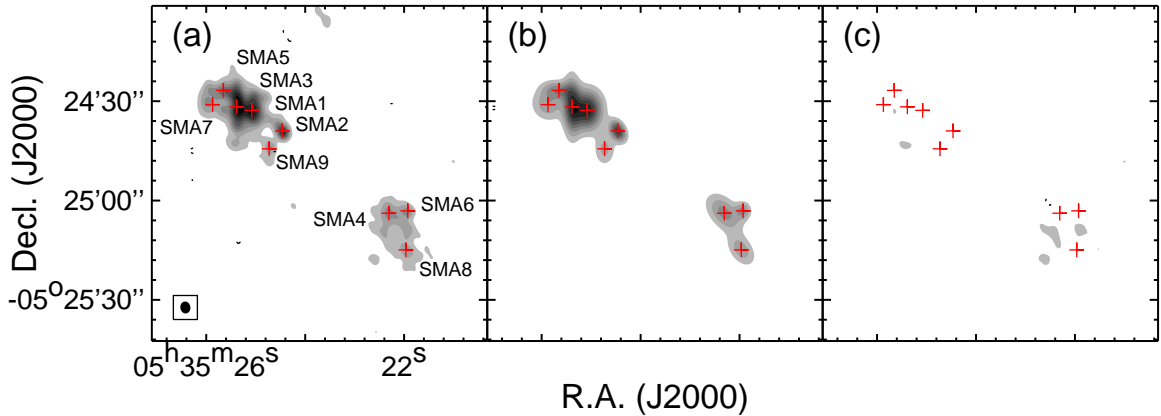}
\caption{(a) Contour map of the 1.2~mm continuum emission observed with the SMA. Contour levels are the same as those in Figure~\ref{fig:spitzer}(b), but shown in filled contours from light to dark gray. Hereafter, an ellipse in the lower left corner shows the SMA synthesized beam accordingly. (b) Contour map of the fitted Gaussian components of the continuum emission, with contour levels the same as those in panel (a). (c) Contour map of the residual emission derived by subtracting the fitted Gaussian components from the observed continuum map, with contour levels the same as those in panel (a). Hereafter, plus symbols, namely SMA1--9, denote the peaks of all the identified dust condensations. 
\label{fig:cont}}
\end{figure}

We identify the dust condensations and measure their fluxes, positions, and sizes from the dust continuum map with a two-step process. First we decompose the continuum map into individual sources with a ``dendrogram'' analysis \citep[e.g.,][]{Goodman09} and a 2D CLUMPFIND algorithm \citep{Williams94}; we find that the results from the two analyses are in general consistent with each other. Second, we perform a multi-component 2D Gaussian fitting on the continuum map using the MIRIAD task IMFIT, with the sources identified by both the dendrogram and CLUMPFIND analyses as the initial estimates. Figure~\ref{fig:cont} shows a comparison between the observed continuum map, the Gaussian fitting results,  and the residual map obtained by subtracting the fitted Gaussian components from the observed map. We measure an RMS noise level of 3.2~mJy\,beam$^{-1}$ from the residual map, which is fairly close to the RMS noise level of the observed map. We identify a total of 9 condensations with peak fluxes at least 7 times the RMS noise level of the observations. These condensations are denoted as SMA1 to SMA9 in a decreasing order of the peak flux in Figure~\ref{fig:cont}(a). SMA2 is a foreground pre-main sequence star with a protoplanetary disk \citep{Mann10}, and will be excluded from the following analyses. 

With the measured flux of each condensation and assuming that the dust emission at 1.2~mm is optically thin, we estimate the dust mass, $M_{\rm dust}$, according to $$M_{\rm dust} = \frac{F_{\nu}D^2}{B_{\nu}(T_{\rm dust})\kappa_{\nu}},$$
where $F_\nu$ is the dust emission flux, $D$ is the distance (417 pc), $B_{v}(T_{\rm dust}$) is the Planck function at dust temperature $T_{\rm dust}$, which is adopted to be 73 K (Section \ref{subsec:lines}) by assuming thermal equilibrium between gas and dust and that all the condensations have the same temperature, and $\kappa_{\nu}$ is the dust opacity adopted to be 1.0~cm$^2$ g$^{-1}$ following \citet{Ossenkopf94} for dust grains with ice mantles in regions of gas densities of order $10^6$ -- $10^8$ cm$^{-3}$. The dust mass is converted to gas mass, $M_{\rm g}$, with a gas-to-dust mass ratio of 100, and the volume density of H$_2$ is computed with the gas mass and the measured size. We list the measured and computed parameters of each condensation in Table~\ref{tab:prop}.
 
\subsection{Molecular Spectral Line Emissions} \label{subsec:lines}
With the two frequency setups we detect spectral line emissions in CS~(5--4), (4--3), C$^{34}$S~(4--3), C$^{33}$S~(4--3), H$_2$CS~(7$_{1,6}$--6$_{1,6}$), (6$_{0,6}$--5$_{0,5}$), (6$_{2,4}$--5$_{2,3}$), and SO~(5$_4$--4$_3$). Among these lines,  maps of CS~(5--4), (4--3), H$_2$CS~(7$_{1,6}$--6$_{1,6}$), and SO~(5$_4$--4$_3$) reveal the distribution of dense molecular gas and complement the dust continuum map. Optically thinner lines of C$^{34}$S and C$^{33}$S~(4--3) are used to estimate the velocity dispersion of the compact structures. The H$_2$CS~(6$_{0,6}$--5$_{0,5}$) and  (6$_{2,4}$--5$_{2,3}$) lines help to constrain the dense gas temperature.

\subsubsection{CS~(5--4), H$_2$CS~(7$_{1,7}$--6$_{1,6}$), and SO~(5$_4$--4$_3$) emissions} \label{subsubsec:distribution}
\begin{figure}
\epsscale{.35}
\plotone{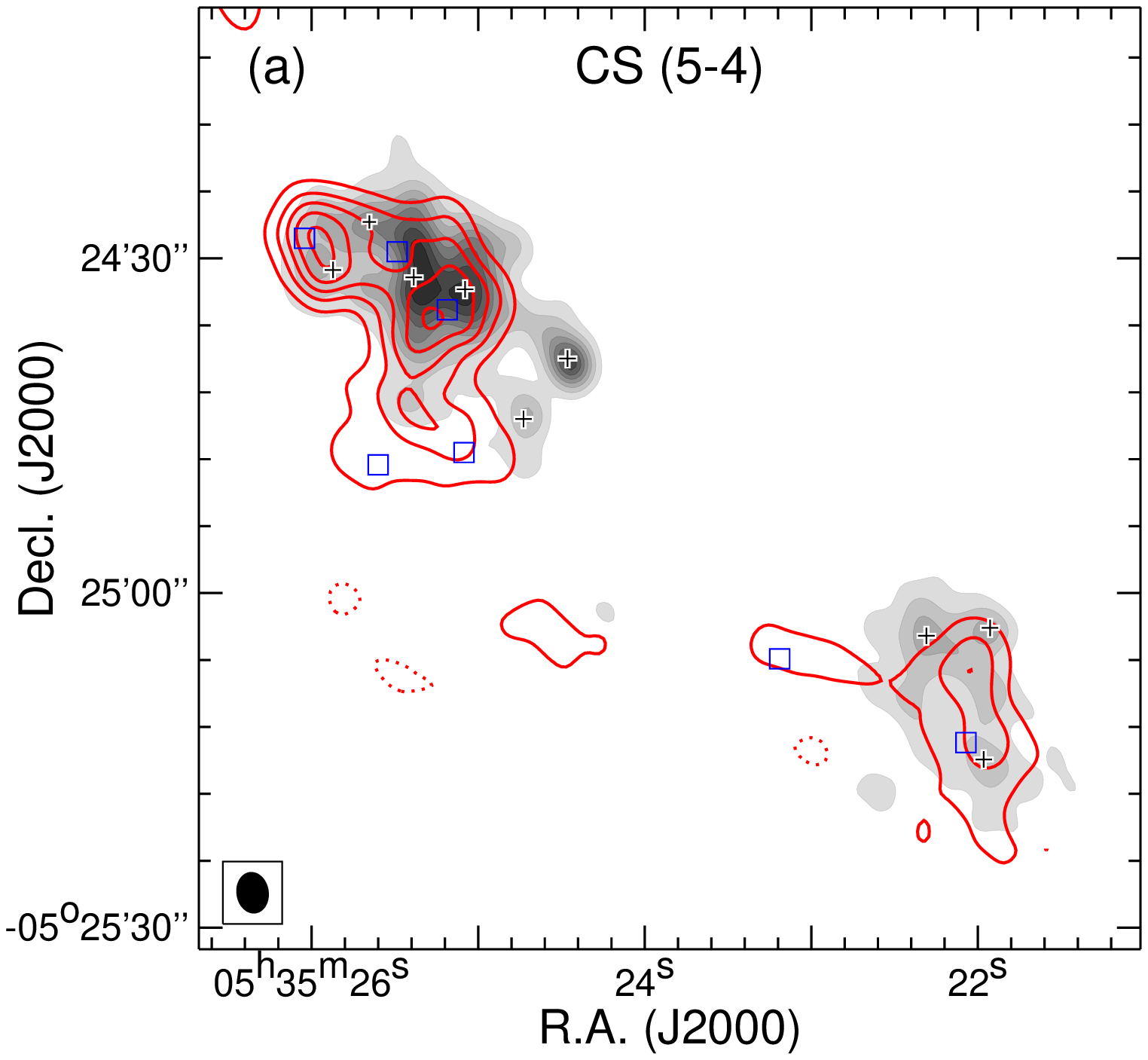}
\plotone{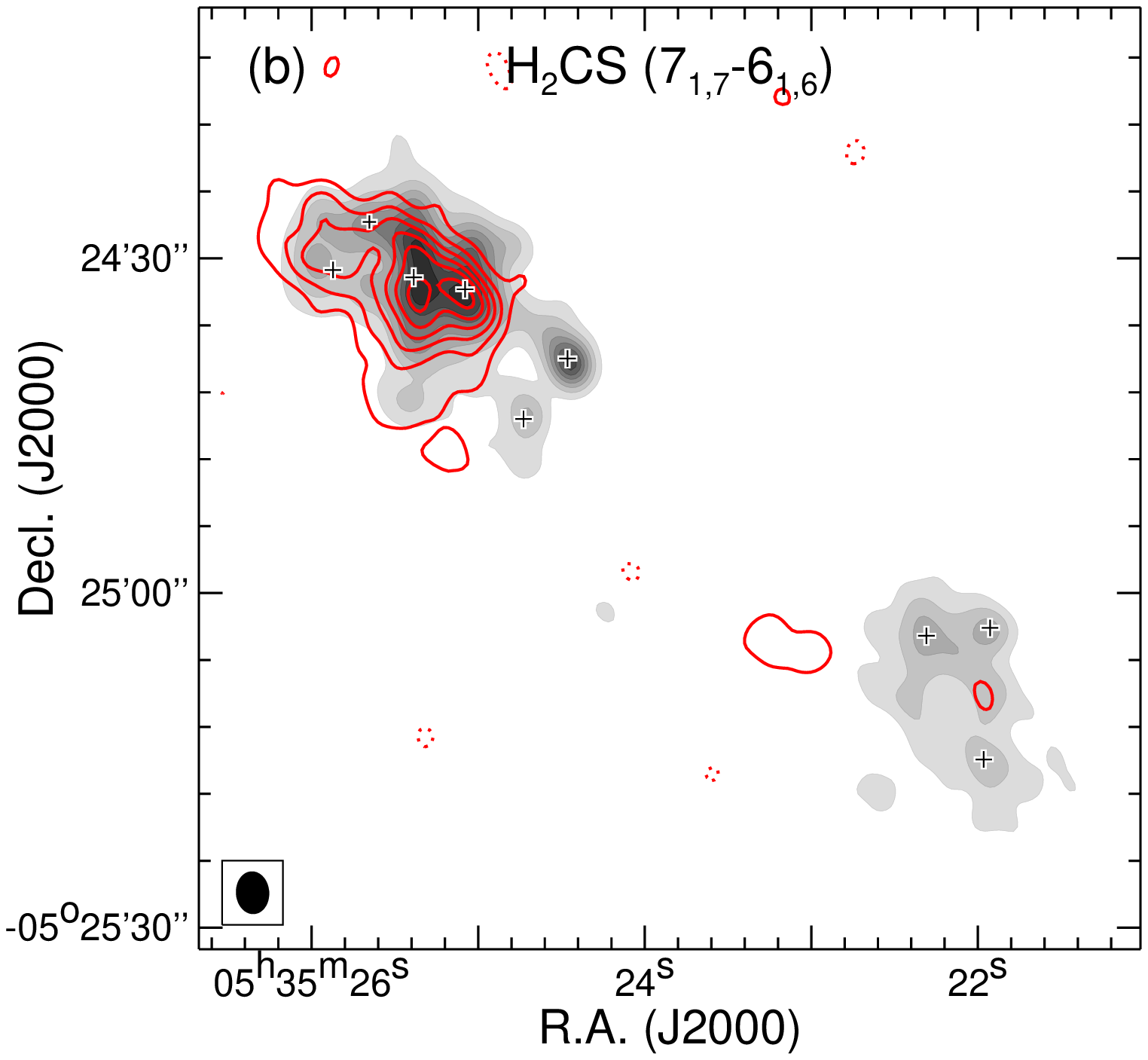}
\plotone{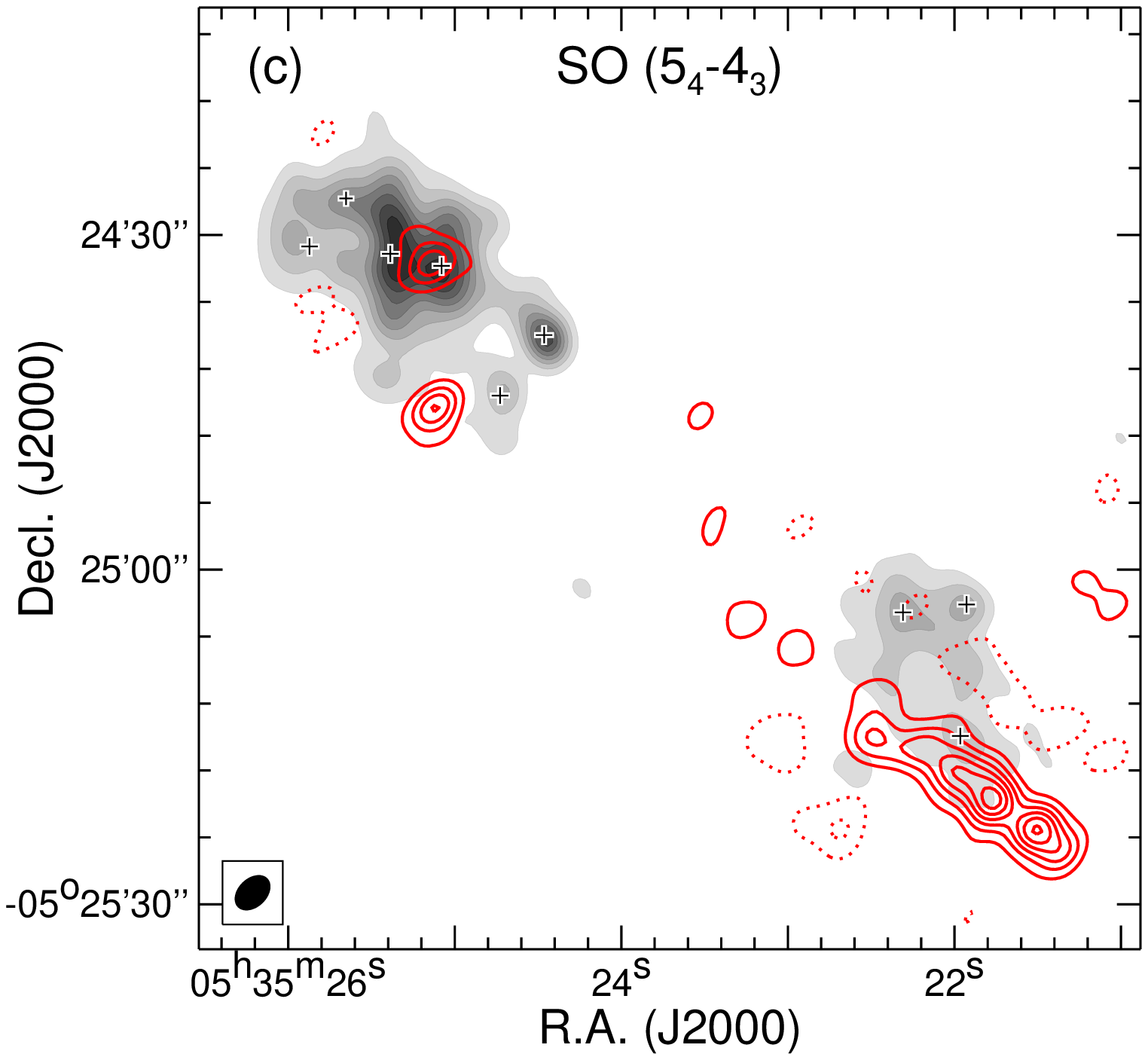}
\caption{(a) Contour map of the velocity integrated emission in CS~(5--4), with contour levels of 15\% to 95\%, by 20\%, of the peak emission of 28.5~Jy\,beam$^{-1}$\,km\,s$^{-1}$. Blue square symbols mark the peak positions of the H$^{13}$CN clumps reported in \citet{Lis03}. (b) Contour map of the velocity integrated emission in H$_2$CS~(7$_{1,7}$--6$_{1,6}$), with contour levels of 12\% to 92\%, by 16\%, of the peak emission of 2.6~Jy\,beam$^{-1}$\,km\,s$^{-1}$. (c) Contour map of the velocity integrated emission in SO~(5$_4$--4$_3$), with contour levels of 15\% to 90\%, by 15\%, of the peak emission of 3.8~Jy\,beam$^{-1}$\,km\,s$^{-1}$. In each panel, filled contours are the same as those in Figure~\ref{fig:cont}(a).
\label{fig:linesm0}}
\end{figure}

\begin{figure}
\plotone{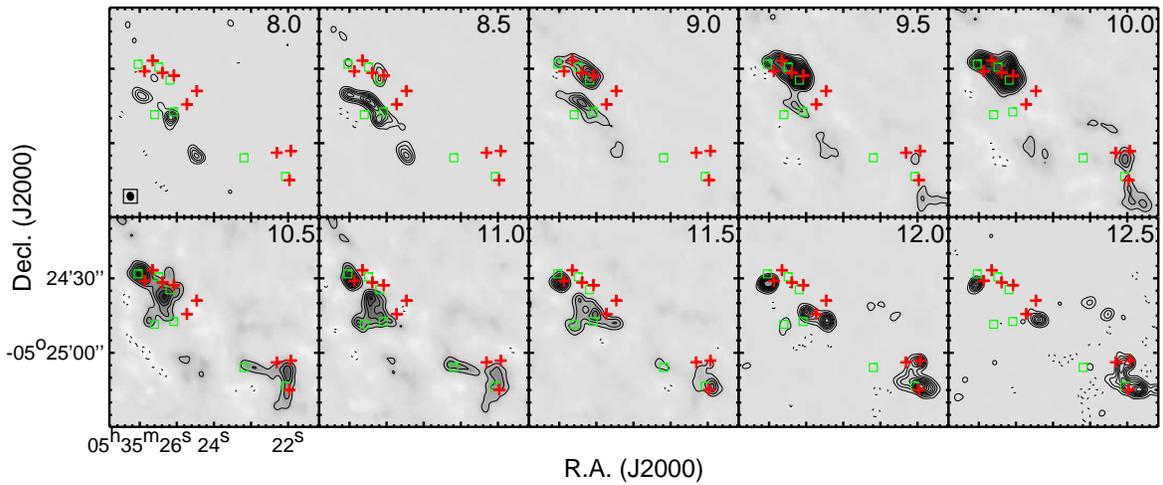}
\caption{Velocity channel maps of the CS~(5--4) emission shown in both grayscale and contours. Starting and stepping contour levels are 0.16, 0.46, 1.54, 1.84, 2.2, 2.8, 2.6, 1.7, 0.46, 0.16~Jy\,beam$^{-1}$ at velocity channels of 8 -- 12.5~km\,s$^{-1}$, respectively. Symbols are the same as those in Figure~\ref{fig:linesm0}(a).
\label{fig:cschan}}
\end{figure}

\begin{figure}
\plotone{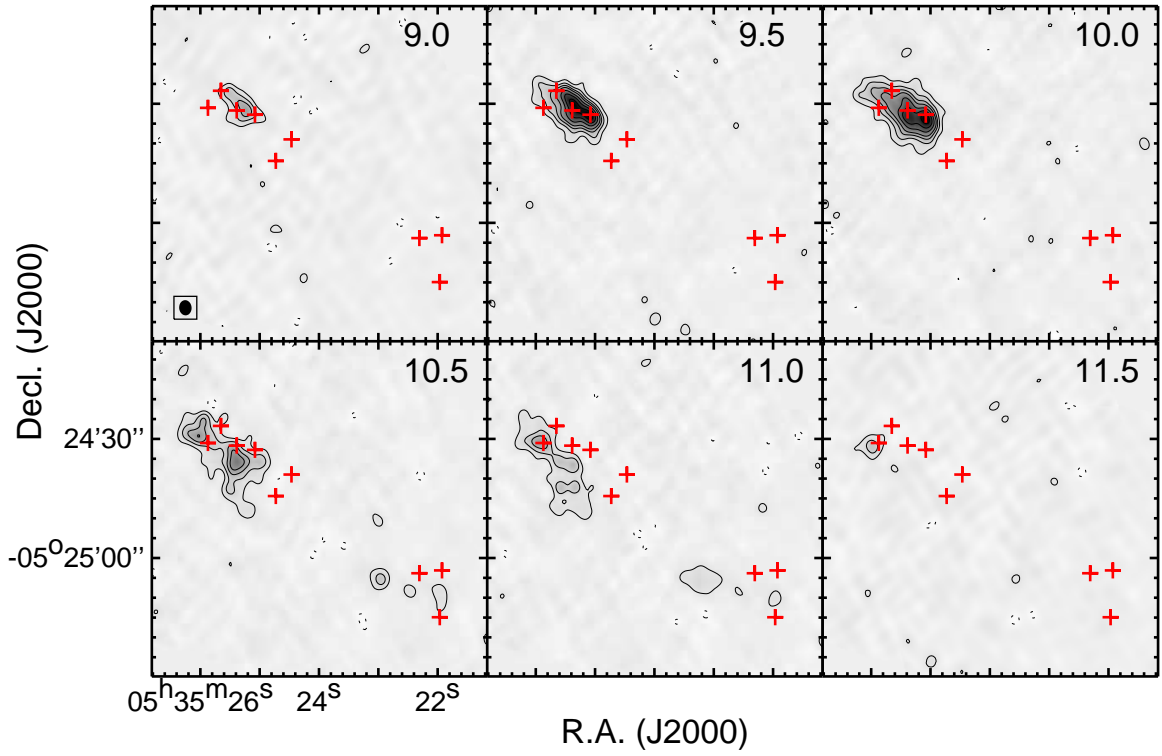}
\caption{Velocity channel maps of the H$_2$CS~(7$_{1,7}$--6$_{1,6}$) emission shown in both grayscale and contours. Contour levels are $(3, 7, 11, 15, 21, 27, 33)\times{\sigma}$, where $\sigma=0.06$~Jy\,beam$^{-1}$.
\label{fig:h2cschan}}
\end{figure}

\begin{figure}
\plotone{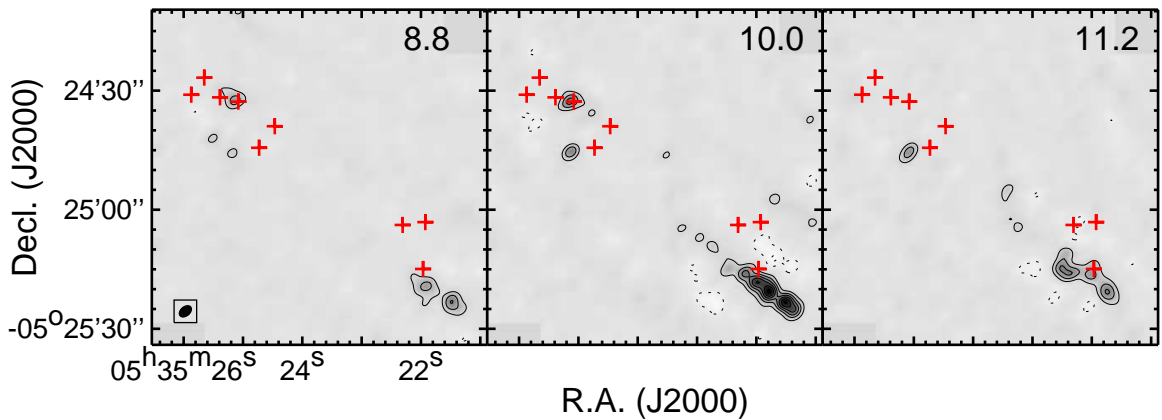}
\caption{Velocity channel maps of the SO~(5$_4$--4$_3$) emission shown in both grayscale and contours. Contours start at and continue in steps of $5\sigma$, where $\sigma=0.06$~Jy\,beam$^{-1}$.
\label{fig:sochan}}
\end{figure}

Figure~\ref{fig:linesm0} shows velocity integrated emissions in CS~(5--4), H$_2$CS~(7$_{1,6}$--6$_{1,6}$), and SO~(5$_4$--4$_3$), and Figures~\ref{fig:cschan}--\ref{fig:sochan} show velocity channel maps of the lines. The CS~(5--4) map in Figure~\ref{fig:linesm0}(a) reveals dense molecular gas around dust condensations SMA1, SMA3, SMA5, and SMA7 in the NE field and SMA6 and SMA8 in the SW field. The CS~(5--4) emission also probes additional structures not detected or very faint in the dust continuum map; such structures are further away from the dissociation front of the Bar, and are seen as a clump to the southeast of SMA9 in Figure~\ref{fig:linesm0}(a) and as clumpy and elongated structures in velocity channels of 8.0--9.0~km\,s$^{-1}$ in Figure \ref{fig:cschan}. The CS~(4--3) map shows gas structures similar to those seen in the CS~(5--4) map, but has a lower sensitivity and image quality, and thus is not shown here. In Figure \ref{fig:linesm0}(b), prominent H$_2$CS~(7$_{1,6}$--6$_{1,6}$) emission is only detected in the NE field, and traces dense gas around SMA1, SMA3, SMA5, and SMA7. A velocity gradient in an orientation perpendicular to the axis of the Bar is seen in the velocity channel map of this line (Figure~\ref{fig:h2cschan}). In general, the brightest emissions in CS~(5--4) and H$_2$CS~(7$_{1,6}$--6$_{1,6}$) are roughly consistent with the dust continuum emission. However, both Figure~\ref{fig:linesm0}(c) and Figure~\ref{fig:sochan} show that the SO~(5$_4$--4$_3$) emission traces dense gas structures very different from those seen in the CS and H$_2$CS emissions. Except a compact structure associated with SMA1 and faint emission associated with SMA8, the SO emission reveals structures not seen in the dust continuum; the emission is dominated by clumpy and elongated structures lying behind the dust condensations in the SW field. 

\subsubsection{H$_2$CS~(6$_{0,6}$--5$_{0,5}$) and (6$_{2,4}$--5$_{2,3}$) emissions} \label{subsubsec:temp}
\begin{figure}
\plotone{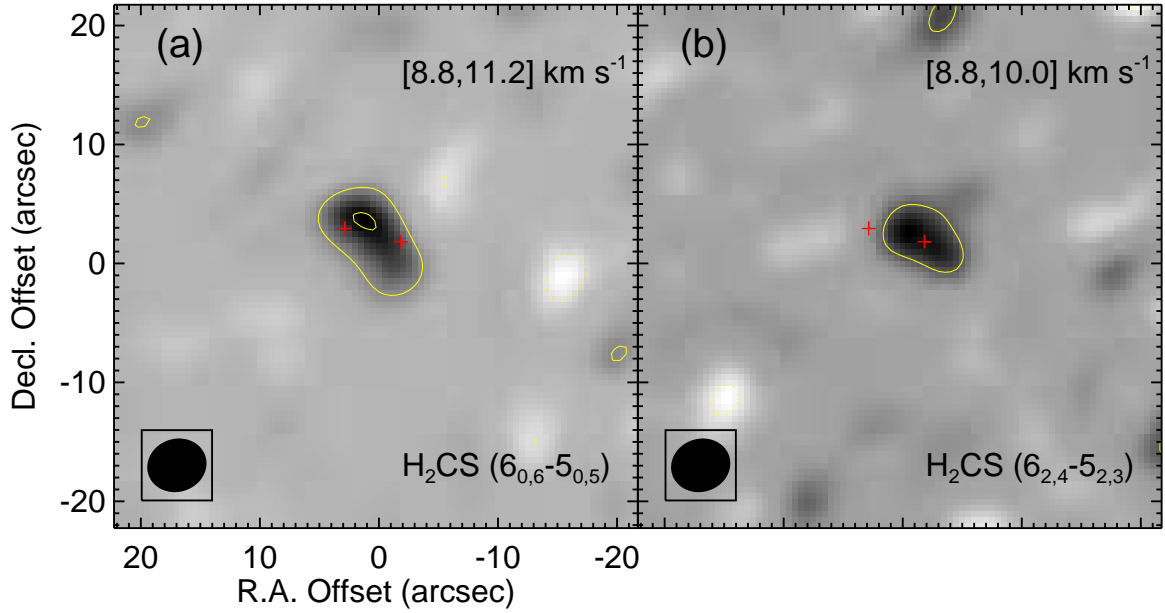}
\caption{(a) Contour map of the H$_2$CS (6$_{0,6}$--$5_{0,5}$) emission integrated from 8.8 to 11.2~km\,s$^{-1}$, lying on top of the grayscale image of the same emission. Starting and spacing contour levels are $3\sigma$, where $\sigma=0.15$~Jy\,beam$^{-1}$ km\,s$^{-1}$. (b) The same as (a), but for the H$_2$CS (6$_{2,4}$--$5_{2,3}$) emission integrated from 8.8 to 10.0~km\,s$^{-1}$, with $\sigma=0.14$~Jy\,beam$^{-1}$ km\,s$^{-1}$. In each panel, two plus symbols mark the peak positions of SMA1 and SMA3, from right to left.
\label{fig:h2cs}}
\end{figure}

H$_2$CS is a slightly asymmetric rotor, a heavier analogue to H$_2$CO \citep{Chandra10}. So one can measure the gas temperature by comparing the intensities of two $K$-components from the same ${\Delta}J=1$ transition of the same symmetry species \citep{Mangum93}. With this motivation we observed H$_2$CS~(6$_{0,6}$--5$_{0,5}$) and (6$_{2,4}$--5$_{2,3}$) lines. We detect H$_2$CS~(6$_{0,6}$--5$_{0,5}$) or (6$_{2,4}$--5$_{2,3}$) emission only toward the NE field. Figure \ref{fig:h2cs} shows velocity integrated emissions of the two lines. Weak emissions are seen around SMA1 and SMA3: the 6$_{0,6}$--5$_{0,5}$ map shows a peak S/N ratio of 6 and the 6$_{2,4}$--5$_{2,3}$ map has a peak S/N ratio of 4. H$_2$CS (6$_{0,6}$--5$_{0,5}$) and (6$_{2,4}$--5$_{2,3}$) have upper energy levels above the ground of 35~K and 87~K, respectively, thus their flux ratio is sensitive to gas temperatures from a few 10 to more than 100~K, suitable for a temperature diagnostic of dense molecular gas in the Bar. We estimate the gas kinematic temperature from the flux ratio toward the peak position of the 6$_{0,6}$--5$_{0,5}$ emission, assuming local thermodynamical equilibrium (LTE) and optically thin emissions for both lines, and obtain a temperature of $73\pm58$~K. This represents an average temperature for gas around SMA1 and SMA3. The large uncertainty of the temperature determination is due to relatively low S/N ratios of the detections.

\subsubsection{C$^{34}$S and C$^{33}$S~(4--3) emissions} \label{subsubsec:dispersion}
\begin{figure}
\plotone{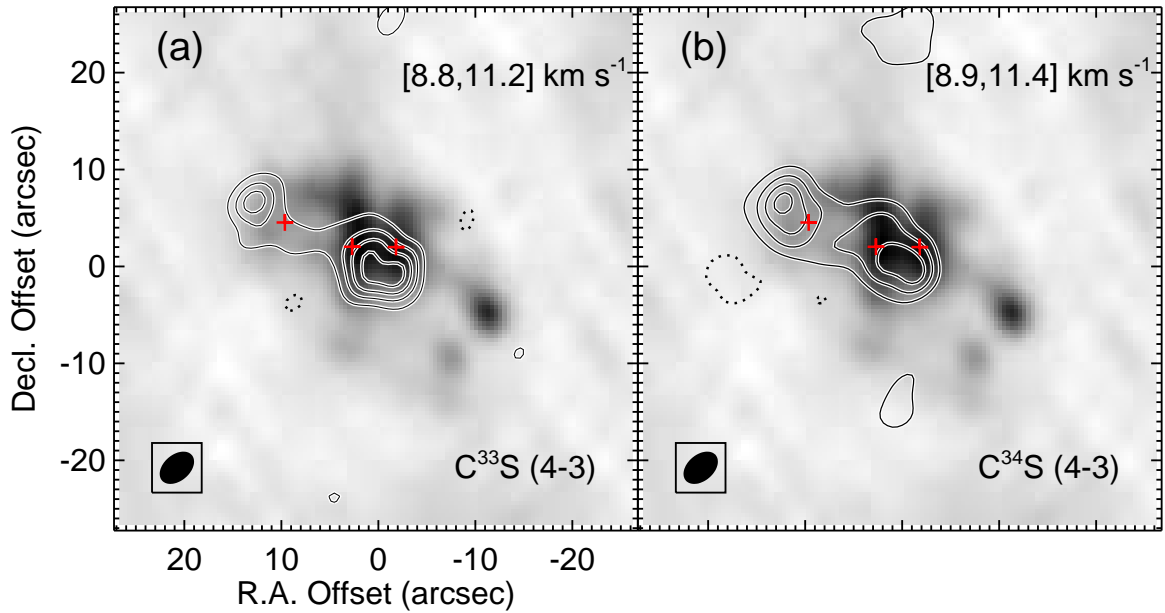}
\caption{(a) Contour map of the C$^{33}$S (4--3) emission integrated from 8.8 to 11.2~km\,s$^{-1}$, lying on top of the grayscale image of the dust emission. Starting and spacing contour levels are $3\sigma$, where $\sigma=0.22$~Jy\,beam$^{-1}$ km\,s$^{-1}$. (b) The same as (a), but for the C$^{34}$S (4--3) emission integrated from 8.9 to 11.4~km\,s$^{-1}$, with $\sigma=0.41$~Jy\,beam$^{-1}$ km\,s$^{-1}$. In each panel, three plus symbols mark the peak positions of SMA1, SMA3, and SMA7, from right to left. 
\label{fig:csm0} }
\end{figure}

\begin{figure}
\epsscale{0.6}
\plotone{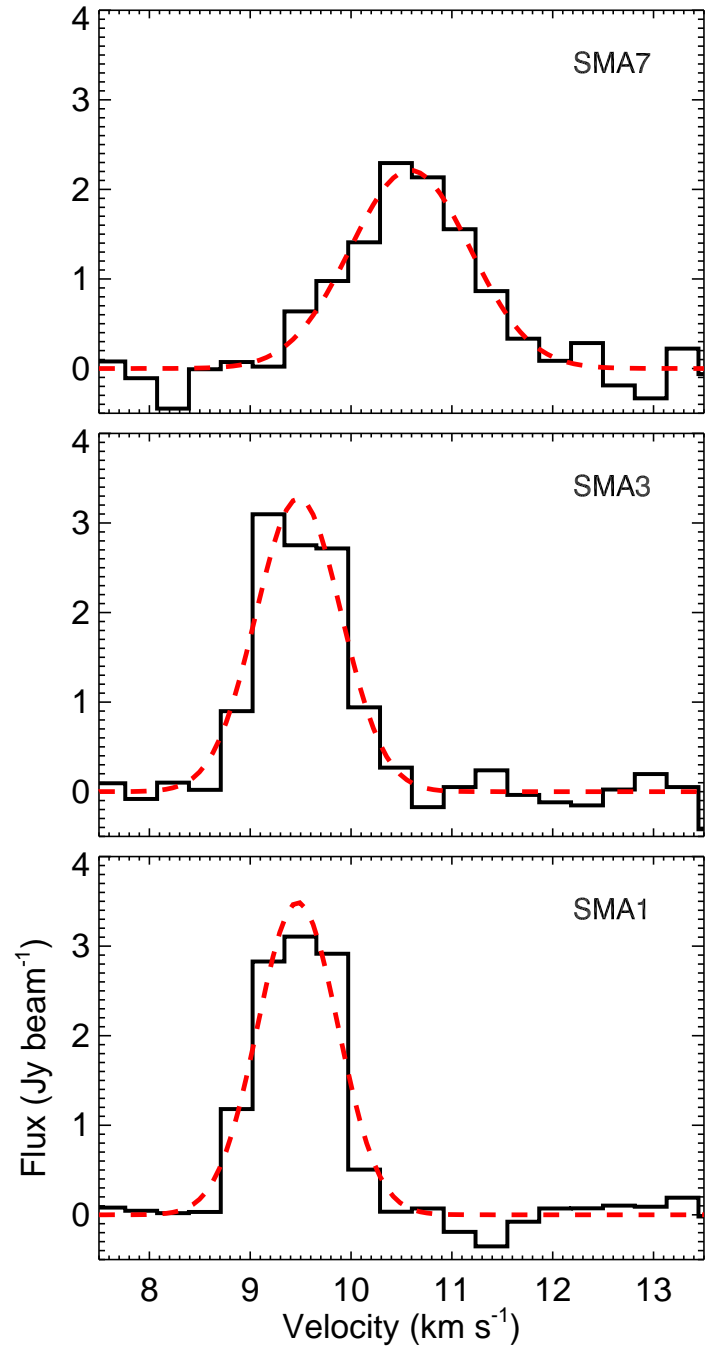}
\caption{C$^{34}$S (4--3) spectra extracted from the peaks of SMA1, SMA3, and SMA7, with observations shown in solid histograms and Gaussian fittings in dashed lines.
\label{fig:csfit} }
\end{figure}

We observed C$^{34}$S~(4--3) with a higher spectral resolution (0.3~km\,s$^{-1}$) to measure the line width toward the dust condensations. The simultaneously observed C$^{33}$S~(4--3) can be used to constrain the optical depth of the C$^{34}$S emission  and improve the accuracy of the line width measurement. Emissions around dust condensations in these two lines are only detected in the NE field. Figure \ref{fig:csm0} shows velocity integrated emissions. Both lines are detected and reveal clumpy structures around SMA1, SMA3, and SMA7. We measure the line widths by performing Gaussian fittings to the C$^{34}$S spectra. Figure~\ref{fig:csfit} shows the C$^{34}$S spectra extracted from the positions of SMA1, SMA3, and SMA7, overlaid with Gaussian fittings. A spectral line could be broadened by the optical depth. Assuming that C$^{34}$S and C$^{33}$S (4--3) have the same excitation temperature, the line ratio is a function of the optical depth and is given by 
$$\frac{F_\nu({\rm C^{34}S})}{F_\nu({\rm C^{33}S})}=\frac{1-e^{-\tau_{\rm C^{34}S}}}{1-e^{-\tau_{\rm C^{34}S}/\chi}},$$
where $F_\nu$ is the peak flux, $\tau_{\rm C^{34}S}$ is the C$^{34}$S optical depth at the line center, and $\chi$ is the abundance ratio of C$^{34}$S to C$^{33}$S and is about 4.9 according to the measured abundance ratio of $^{34}$S to $^{33}$S in the Orion KL region \citep{Persson07}. After smoothing the C$^{34}$S spectra to a spectral resolution of 1.2~km\,s$^{-1}$, the same as the C$^{33}$S spectra, we derive optical depths of $2.29\pm0.24$, $2.86\pm0.27$, and $0.70\pm0.29$ toward SMA1, SMA3, and SMA7, respectively, for the C$^{34}$S lines. The intrinsic line width, ${\Delta}V_{\rm int}$, is then deduced following  \citet{Beltran05}: $$\frac{{\Delta}V_{\rm obs}}{{\Delta}V_{\rm int}} = \frac{1}{\sqrt{ln2}}\sqrt{-ln\left[-\frac{1}{\tau_{\rm C^{34}S}}ln\left(\frac{1+e^{-\tau_{\rm C^{34}S}}}{2}\right)\right]},$$ where ${\Delta}V_{\rm obs}$ is the line width derived from Gaussian fittings (see Table \ref{tab:prop}). Consequently, the non-thermal velocity dispersion, $\sigma_{\rm nth}$, can be calculated from $\sigma_{\rm nth}=\sqrt{{\Delta}V_{\rm int}^2/8ln2-\sigma_{\rm th}^2}$, where $\sigma_{\rm th}=0.12$~km\,s$^{-1}$ is the thermal velocity dispersion at 73~K. We find that $\sigma_{\rm nth}$ ranges from 0.24 to 0.50~km\,s$^{-1}$ (Table \ref{tab:prop}), and is smaller than or at most comparable with the speed of sound, $c_{\rm s}$, which is 0.51~km\,s$^{-1}$ at 73~K.

\section{Discussion} \label{sec:discuss}
\subsection{Dense molecular gas revealed by the dust continuum and molecular spectral line emissions} \label{subsec:densegas}
We detect clumpy and dense gas structures in both dust continuum and molecular spectral line observations of the Orion Bar. By comparing with the IRAC image, we find that the dust condensations lying from immediately behind the dissociation front to an extension of $\sim15''$ (0.03~pc) toward the interior of the molecular cloud behind the Bar. Among the detected spectral lines, C$^{34}$S, C$^{33}$S~(4--3) and H$_2$CS~(6$_{0,6}$--5$_{0,5}$), (6$_{2,4}$--5$_{2,3}$) show unresolved and faint emissions toward the brightest dust condensations. The other spectral lines reveal a larger extent of dense gas and could complement the dust continuum map. Most CS~(5--4) clumps and all the H$_2$CS~(7$_{1,7}$--6$_{1,6}$) clumps appear to be associated with the dust emission. The CS emission also reveals gas structures further away from the dissociation front. We then compare previous PdBI H$^{13}$CN~(1--0) observations with our spectral line maps for a more comprehensive overview of the distribution of dense molecular gas. In Figure~\ref{fig:linesm0}(a), all the H$^{13}$CN clumps seem to be associated with the CS emission. Such an association is much clearer in Figure~\ref{fig:cschan}, where a CS counterpart of every H$^{13}$CN clump can be identified in some certain velocity channels. The SO~(5$_4$--4$_3$) map, however, has a distinctive behavior; the emission in this line is dominated by structures devoid of dust emission and undetected in any other line. 

Therefore, while the dust emission reveals a majority of the dense gas structures within a distance of $\sim$0.03~pc from the dissociation front, some molecular spectral lines such as CS, H$^{13}$CN, and SO lines could trace additional structures deeper into the cold molecular cloud. In addition to the variation of physical conditions as a function of distance from the dissociation front, chemistry should also play a role in producing the difference in the distribution of various spectral line emissions. We refrain from discussing chemical stratification in the distribution of molecular gas in the Bar, since it is out of the scope of this work. Here we focus on the dust condensations, which are characteristic of dense gas structures immediately behind the dissociation front, and their masses could be estimated based on the dust emission.

\subsection{On the nature of the detected dust condensations} \label{subsec:diffmass}
We identify a total of 9 condensations from the 1.2~mm continuum map. Except one condensation (SMA2) associated with a protoplanetary disk and regarded as a foreground source, all the other condensations are dense gas structures in the Bar. The masses of the condensations are estimated to be $\sim$0.03 -- 0.3 $M_{\odot}$, a few times lower than the virial masses of the H$^{13}$CN clumps \citep{Lis03} and one to two orders of magnitude greater than that of the HCO$^+$ substructures \citep{Goicoechea16}. Are these condensations gravitationally bound? Would they collapse to form stars? How were they formed?

To address these questions, we first calculate the Jeans mass, $M_{\rm J}$, of the condensations following $$M_{\rm J} = \frac{4\pi}{3}(\frac{\lambda_{\rm J}}{2})^3\rho,$$ where $\lambda_{\rm J} = \sqrt{{\pi}c_{\rm s}^2/G\rho}$ is the Jeans length, and $\rho$ is the mass density. By comparing $M_{\rm J}$ with $M_{\rm g}$ listed in Table \ref{tab:prop}, it is immediately clear that all the detected condensations have masses more than an order of magnitude lower than $M_{\rm J}$. For three condensations, SMA1, SMA3, and SMA7, with the measured line widths we can calculate the virial mass, $M_{\rm vir}$, following $$M_{\rm vir} = 1165(\frac{R}{\rm 1~pc})(\frac{\sigma_{\rm tot}}{\rm 1~km~s^{-1}})^{2},$$
where $R$ is taken as half of the geometric mean of the deconvolved major and minor axes of the condensation (Table \ref{tab:prop}), and $\sigma_{\rm tot} = \sqrt{\sigma_{\rm nth}^2 + c_{\rm s}^2}$ is the total velocity dispersion. From Table \ref{tab:prop}, the calculated virial mass is close to the Jean mass, and significantly greater than the gas mass of the condensations. The condensations in the Bar are believed to be surrounded by a less dense medium and the external pressure may play a role in the dynamical evolution of the condensations. We then estimate the Bonnor-Ebert mass, $M_{\rm BE}$, given by $1.182c_{\rm s}^3/\sqrt{G^3P_0}$, where $P_0$ is the external pressure \citep{McKee07}. The density of the surrounding medium is suggested to be $10^4$ -- $10^5$~cm$^{-3}$ \citep{Hogerheijde95,Young00,Goicoechea16}, resulting in $M_{\rm BE}\sim22$ -- 7.1~$M_{\odot}$, which is again orders of magnitude greater than $M_{\rm g}$. All these calculations indicate that the detected dust condensations have masses significantly lower than that is required for making them gravitationally bound or for self-gravity to play a crucial role in their dynamical evolution. Thus it seems unlikely that the dust condensations could collapse to form stars in the future. 

\begin{figure}
\epsscale{.6}
\plotone{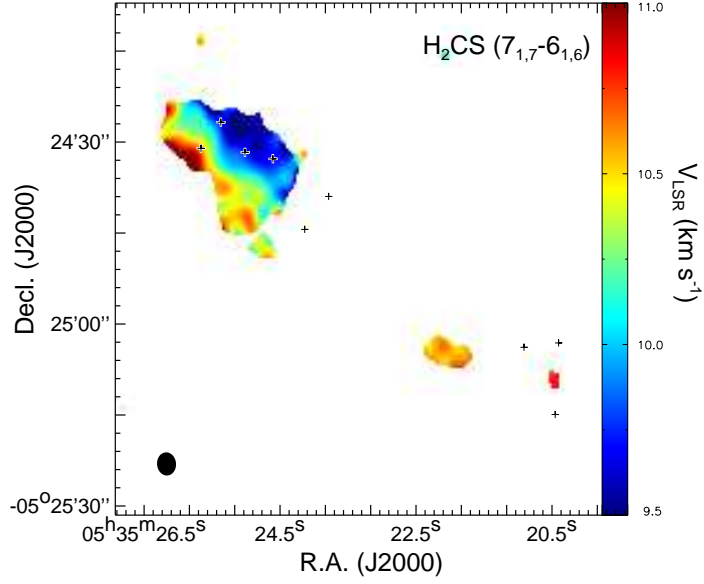}
\caption{Colored image of the first moment (intensity-weighted velocity) map of the H$_2$CS~(7$_{1,7}$--6$_{1,6}$) emission. A vertical color bar to the right indicates the velocity scale shown in the image. 
\label{fig:h2csm1} }
\end{figure}

The dust condensations could not arise from a thermal Jeans fragmentation process. If that is the case, with a density of $10^4$ -- $10^5$~cm$^{-3}$ for the surrounding medium, one may expect the mass of the condensations on the order of 20--50~$M_{\odot}$ (the Jeans mass at $10^5$ -- $10^4$~cm$^{-3}$) and the nearest separations between the condensations around 0.2--0.5~pc (the Jeans length at $10^5$ -- $10^4$~cm$^{-3}$); both are clearly inconsistent with the observations. Alternatively, small dense structures can be temporary density fluctuations frequently created and destroyed by supersonic turbulence \citep[e.g.,][]{Elmegreen99,Biskamp03,Falgarone03}. However, $\sigma_{\rm nth}$ is found to be subsonic or at most transonic. \citet{Goicoechea16} also found that there is only a gentle level of turbulence in the Bar. So the turbulence dose not seem to be strong enough to produce the condensations. Another force that could potentially compress the cloud and produce high density structures is a high pressure wave from the expansion of the H{\scriptsize II} region. \citet{Goicoechea16} detected a fragmented ridge of high density substructures at the molecular cloud surface and three periodic emission maxima in HCO$^+$ (4--3) from the cloud edge to the interior of the Bar, providing evidence that a high pressure wave has compressed the cloud surface and moved into the cloud to a distance of $\sim$15$''$ from the dissociation front. The dust condensations are also located within a distance of $15''$ from the dissociation front, and thus are very likely over-dense structures created as the compressive wave passed by. The complex clumpy appearance of the condensations is probably related to the front instability of the compressive wave \citep{Goicoechea16}, or an instability developed across different layers of the Bar \citep[e.g., the thin-shell instability,][]{Garcia96}. The velocity structure of dense gas around the dust condensations may provide insights into the feasibly of this scenario. Figure~\ref{fig:h2csm1} shows the intensity-weighted velocity map of the H$_2$CS~(7$_{1,7}$--6$_{1,6}$) emission. A velocity gradient in a northwest-southeast direction (i.e., along the direction of the propagation of the compressive wave) is seen in the map, and is consistent with the scenario that the gas is being compressed by a high pressure wave.

\subsection{Uncertainty of the mass estimate} \label{subsec:uncert}
We estimate the masses of the condensations based on the dust emission. The uncertainty of this estimate mainly depends on the accuracy in determining the dust opacity and temperature. We adopted a dust opacity of 1 cm$^2$ g$^{-1}$ (at 249~GHz), which is equivalent to adopting a dust opacity index, $\beta$, of 1.5 according to $\kappa_{\nu}=10(\nu/1.2{\rm THz})^{\beta}$ \citep{Hildebrand83}. It agrees well with the value measured by \citet{Arab12}, who obtained $\beta=1.62\pm0.13$ toward the dust emission peak inside the Bar based on Herschel 70 to 500 $\mu$m observations. This renders support for the appropriateness of our adopted dust opacity. Hence we expect that the largest uncertainty in the mass estimate is ascribed to the dust temperature determination. 

With the newly detected H$_2$CS lines, we measured a gas temperature of $73\pm58$ K. There have been a number of molecular spectral line observations providing constraints on the temperature of dense gas in the Bar \citep[e.g.,][]{Hogerheijde95,Batrla03,Goicoechea11,Goicoechea16,Nagy17}. These studies suggest a gas temperature of $\gtrsim$100~K; it appears that our derived value of 73~K lies at the lower end of the range. However, the dust temperature could be lower than the gas temperature. \citet{Goicoechea16} illustrate that for the gas behind the dissociation front in a PDR such as the Orion Bar, the gas temperature decreases from 500 K to 100 K while the dust temperature varies from 75 to 50 K. Indeed, \citet{Arab12} measured a dust temperature of 49~K toward the dust emission peak inside the Bar based on Herschel observations. Considering that the dust condensations detected in our SMA map are located close to the dissociation front, a value of 50~K may represent a lower limit for the dust temperature of the condensations. For $T_{\rm dust}\sim50$~K,  $M_{\rm g}$ in Table \ref{tab:prop} would be increased by about 50\% and $M_{\rm J}$ would be decreased by about 50\%. But $M_{\rm J}$ and $M_{\rm vir}$ would be still clearly greater than $M_{\rm g}$, indicating that the condensations are anyway gravitationally unbound. In the less likely case that $T_{\rm dust}>73$~K, $M_{\rm J}$ becomes greater and $M_{\rm g}$ gets lower, and all the discussions remain unchanged.

\section{Summary} \label{sec:sum}
We have performed SMA dust continuum and molecular spectral line observations toward the Orion Bar. The 1.2~mm continuum map reveals, for the first time, dust condensations in the Bar at arc~second resolutions. These condensations are distributed immediately behind the dissociation front. They have an average temperature of about 73~K, and exhibit subsonic to transonic turbulence. The masses of the condensations span a range of $\sim$0.03--0.3~$M_{\odot}$, which are all too low to enable self-gravity to play a crucial role. Thus the condensations do not seem to be able to collapse and form stars. The formation of these condensations is mostly likely induced by a compressive wave originating from the expansion of the H{\scriptsize II} region.

\acknowledgments K.Q. acknowledges the support from National Natural Science Foundation of China (NSFC) through grants 11473011 and 11590781. Q. Z. acknowledges the support from NSFC through grant 11629302.

\clearpage

\floattable
\begin{deluxetable}{lccccccccccc}
\tablenum{1}
\tablecaption{Measured and computed parameters of dust condensations} \label{tab:prop}
\tablewidth{0pt}
\tablehead{
 \colhead{Source} & \multicolumn2c{Peak Position} & \colhead{Size} & \colhead{Peak Flux}  & \colhead{Integrated Flux} & \colhead{$M_{\rm g}$} & \colhead{$n_{\rm H_2}$} & \colhead{$M_{\rm J}$} & \colhead{${\Delta}V_{\rm obs}$} & \colhead{${\sigma}_{\rm nth}$} & \colhead{$M_{\rm vir}$} \\
   & $\alpha_{\rm J2000}$ & $\delta_{\rm J2000}$ & $('')\times('')$ & (Jy\,beam$^{-1}$) & (Jy) & $(M_{\odot})$ & (10$^6$cm$^{-3}$) & $(M_{\odot})$ & (km\,s$^{-1}$) & (km\,s$^{-1}$) & $(M_{\odot})$
}
\decimalcolnumbers
\startdata
SMA1 & 5:35:25.08   & $-5$:24:32.8 & $8.4\times6.3$   & 0.069 & 0.40   & 0.28 & 2.8   & 3.2 & 0.89  & 0.24 & 2.7  \\
SMA2 & 5:35:24.46   & $-5$:24:39.0 & $3.8\times1.5$   & 0.060 & 0.10   &         &         &       &          &         &        \\
SMA3 & 5:35:25.39   & $-5$:24:31.7 & $13.0\times2.7$ & 0.049 & 0.24   & 0.17 & 3.2   & 3.0 & 0.95  & 0.25 & 2.3  \\
SMA4 & 5:35:22.31   & $-5$:25:03.8 & $9.6\times5.3$   & 0.031 & 0.18   & 0.13 & 1.3   & 4.5 &          &         &        \\
SMA5 & 5:35:25.65   & $-5$:24:26.7 & $6.8\times2.5$   & 0.030 & 0.09   & 0.06 & 1.9   & 3.7 &          &         &        \\
SMA6 & 5:35:21.93   & $-5$:25:03.1 & $2.2\times2.1$   & 0.027 & 0.04   & 0.03 & 16.9 & 1.4 &          &         &        \\
SMA7 & 5:35:25.87   & $-5$:24:31.0 & $6.9\times6.4$   & 0.026 & 0.13   & 0.09 & 0.9   & 5.4 & 1.37  & 0.50 & 4.0  \\
SMA8 & 5:35:21.97   & $-5$:25:14.9 & $7.5\times4.2$   & 0.024 & 0.10   & 0.07 & 1.5   & 4.2 &          &         &        \\
SMA9 & 5:35:24.73   & $-5$:24:44.4 & $6.7\times3.8$   & 0.021 & 0.07   & 0.05 & 1.3   & 4.5 &          &         &        \\
\enddata
\tablecomments{Entries in col. (2--6) are parameters derived from 2D Gaussian fittings. The number density given in col. (8) is calculated as $n_{\rm H_2}=3M_{\rm g}/({\mu_{\rm H_2}}m_{\rm H}4{\pi}R^{3}) $, where $R$ is half of the geometric mean of the deconvolved major and minor axes, $m_{\rm H}$ is the proton mass, and $\mu_{\rm H_2} = 2.8$ is the mean molecular weight of H$_2$ \citep{Kauffmann08}. The line width given in col. (10) has been corrected for a spectral resolution of 0.3~km\,s$^{-1}$.
} 
\end{deluxetable}


\begin{thebibliography}{}
\bibitem[Andree-Labsch et al. (2017)]{Andree17}Andree-Labsch, S., Ossenkopf-Okada, V., \& R\"ollig, M. 2017, \aap, 598, A2.
\bibitem[Ag\'undez et al. (2008)]{Agundez08}Ag\'undez, M., Cernicharo, J., \& Goicoechea, J. R. 2008, \aap, 483, 831
\bibitem[Arab et al. (2012)]{Arab12}Arab, H., Abergel, A., Habart, E. et al. 2012, \aap, 541, A19
\bibitem[Batrla \& Wilson (2003)]{Batrla03}Batrla, W. \& Wilson, T. L. 2003, \aap, 408, 231 
\bibitem[Beltr\'an et al. (2005)]{Beltran05}Beltr\'an, M. T., Cesaroni, R., Neri, R. et al. 2005, \aap, 435, 901
\bibitem[Biskamp (2003)]{Biskamp03}Biskamp, D. 2003, Magnetohydrodynamic turbulence (1st ed.; Cambridge: Cambridge Univ. Press)
\bibitem[Chandra et al. (2010)]{Chandra10}Chandra, S., Kumar, A., \& Sharma, M. K. 2010, New Astronomy, 15, 318
\bibitem[Elmegreen (1999)]{Elmegreen99}Elmegreen, B. G. 1999, in The Physics and Chemistry of the Interstellar Medium, ed. V. Ossenkopf, J. Stutzki, \& G. Winnewisser (Herdeke: CGA), 77
\bibitem[Falgarone \& Passot (2003)]{Falgarone03}Falgarone, E. \& Passot, T. 2003, Turbulence and Magnetic fields in Astrophysics (Berlin: Springer)
\bibitem[Fuente et al. (2008)]{Fuente08}Fuente, A., García-Burillo, S., Usero, A., Gerin, M., Neri, R., et al. 2008, \aap, 492, 675
\bibitem[Garc\'{i}a-Segura \& Franco (1996)]{Garcia96}Garc\'{i}a-Segura, G. \& Franco, J. 1996, \apj, 469, 171
\bibitem[Goicoechea et al. (2011)]{Goicoechea11}Goicoechea, J. R., Joblin, C., Contursi, A. et al. 2011, \aap, 530, L16
\bibitem[Goicoechea et al. (2016)]{Goicoechea16}Goicoechea, J. R., Pety, J., Cuadrado, S. et al. 2016, \nat, 537, 207
\bibitem[Goodman et al. (2009)]{Goodman09}Goodman, A., Rosolowsky, E. W., Borkin, M. A. et al. 2009, \nat, 457, 63
\bibitem[Gorti \& Hollenbach (2002)]{Gorti02}Gorti, U., \& Hollenbach, D. 2002, \apj, 573, 215
\bibitem[Hildebrand (1983)]{Hildebrand83}Hildebrand, R. H. 1983, QJRAS, 24, 267
\bibitem[Hogerheijde et al. (1995)]{Hogerheijde95}Hogerheijde, M. R., Jansen, D. J., \& van Dishoeck, E. F. 1995, \aap, 294, 792
\bibitem[Hollenbach \& Tielens (1997)]{Hollenbach97}Hollenbach, D. J. \& Tielens, A. G. G. M. 1997, AR\aap, 35, 179
\bibitem[Jansen et al. (1995)]{Jansen95}Jansen, D. J., Spaans, M., Hogerheijde, M. R., \& van Dishoeck, E. F. 1995, \aap, 303, 541
\bibitem[Kauffmann et al. (2008)]{Kauffmann08}Kauffmann, J., Bertoldi, F., Bourke, T. L., Evans II, N. J., \& Lee, C. W. 2008, \aap, 487, 993
\bibitem[Leurini et al. (2010)]{Leurini10}Leurini, S., Parise, B., Schilke, P., Pety, J., \& Rolffs, R. 2010, \aap, 511, A82.
\bibitem[Leurini et al. (2006)]{Leurini06}Leurini, S., Rolffs, R., Thorwirth, S. et al. 2006, \aap, 454, L47
\bibitem[Lis \& Schilke (2003)]{Lis03}Lis, D. C. \& Schilke, P. 2003, \apjl, 597, L145
\bibitem[MaKee \& Ostriker (2007)]{McKee07}McKee, C. F. \& Ostriker, E. C. 2007, ARA\&A, 45, 565
\bibitem[Mangum \& Wootten (1993)]{Mangum93}Mangum, J. G. \& Wootten, A. 1993, \apjs, 89, 123
\bibitem[Mann \& Williams (2010)]{Mann10}Mann, R. K., \& Williams, J. P. 2010, \apj, 725(1), 430
\bibitem[Menten et al. (2007)]{Menten07}Menten, K. M., Reid, M. J., Forbrich, J., \& Brunthaler, A. 2007, \aap, 474, 515
\bibitem[Nagy et al. (2017)]{Nagy17}Nagy, Z., Choi, Y., Ossenkopf-Okada, V. et al. 2017, \aap, 599, A22
\bibitem[Ossenkopf \& Henning (1994)]{Ossenkopf94}Ossenkopf, V. \& Henning, T. 1994, \aap, 291, 943
\bibitem[Peng et al. (2012)]{Peng12}Peng, T.-C., Wyrowski, F., Zapata, L. A., G\"{u}sten, R., \& Menten, K. M. 2012, \aap, 538, A12 
\bibitem[Persson et al. (2007)]{Persson07}Persson, C. M., Olofsson, A. O. H., Koning, N. et al. 2007, \aap, 476, 807
\bibitem[Tielens et al. (1993)]{Tielens93}Tielens, A. G. G. M., Meixner, M. M., van der Werf, P. P. et al. 1993, Science, 262, 86
\bibitem[van der Wiel et al. (2009)]{van09}van der Wiel, M. H. D., van der Tak, F. F. S., Ossenkopf, V. et al. 2009, \aap, 498, 161
\bibitem[Walmsley et al. (2000)]{Walmsley00}Walmsley, C. M., Natta, A., Oliva, E., \& Testi, L. 2000, \aap, 364, 301
\bibitem[Williams et al. (1994)]{Williams94}Williams, J. P., De Geus, E. J., \& Blitz, L. 1994, \apj, 428, 693
\bibitem[Young Owl et al. (2000)]{Young00}Young Owl, R. C., Meixner, M. M., Wolfire, M., Tielens, A. G. G. M., \& Tauber, J. 2000, ApJ, 540, 886

\end{thebibliography}
\end{document}